\documentclass[11pt]{article}

\usepackage[utf8]{inputenc}
\usepackage[T2A]{fontenc}
\usepackage[english]{babel}
\usepackage{xcolor,amsmath,amssymb}
\usepackage {hyperref,indentfirst, cite}

\title{Problems of the correspondence principle for the recombination cross section in dark plasma}
\author{Belotsky Konstantin\\k-belotsky@yandex.ru\\
	Esipova Ekaterina\\esipovaea@gmail.com\\
	Kalashnikov Dmitriy\\impermast@gmail.com\\
    Letunov Andrei\\letunovandrey11@yandex.ru\\
    	National Research Nuclear University MEPhI, 115409 Moscow, Russia}

\begin{document}
	\maketitle
	
	\begin{abstract}
We raise the issues concerning correspondence principle in  description of a recombination of oppositely charged particles. These issues have come from cosmological dark matter (DM) problem investigations. 
Particles possessing Coulomb-like interaction are considered. 
Such Coulomb-like interaction between DM particles is assumed though the problem seems to be more general.
Analysis showed that usage of different semiclassical approaches leads to the apparent discrepancy between numbers of recombination acts. 
We attempted to find some conditions under which classical cross-section (which relates to multiple soft photon process) reduces to quantum one, which is obtained in semi-classical approximation (Kramers' formula).
We just draw attention to this and provide some (not decisive) arguments.
	\end{abstract}
	
	\noindent Keywords: {correspondence principle, dark matter, dark plasma, collision theory, semiclassical approach} 
\date{July 2020}

\maketitle

\section{Introduction}\label{s1}

Investigation of dark matter (DM) is one of the most important problems in cosmology and particle physics. Many experiments are being carried out to detect DM particles and explore its properties \cite{bernabei2008first,cdms2010dark,aprile2005xenon,abe2008xmass, xenoncollaboration2020projected}. 
A part of the models considers self-interacting DM 
including Coulomb-like interaction \cite{petraki2014self, cyr2013cosmology,von2014bound,belotsky2006composite,Feldman_2010,Tulin_2018,Kamada_2017,Kajiyama_2013}. Within the framework of such models, several disadvantages of the standard $\Lambda$CDM scenario can be avoided.
These are cuspy density profile of the halo, number of small halos and similar other.

Models with Coulomb-like interaction lead to 
the differences in cosmological evolution. Dark charged particles can form a bound state. 
If they are a particle and an antiparticle, they
annihilate, what significantly reduces their density. 
If they are not particle and antiparticle but have opposite "dark" charges (they can be called in this case "dark electron" and "dark proton"), binding processes (recombination) lead to a decrease of free dark charged particles. This significantly changes dynamics of dark matter during the formation of structures in the Universe \cite{von2014bound,Belotsky_2016} and thermodynamical evolution which we considered previously \cite{Esipova_temperatureevolution}. Thus, accounting for recombination is an essential part of such models.

Theory of atomic kinetics is deeply studied section of plasma physics \cite{vajnshtejn19791,bates2012atomic}. Density evolution of specific particle's sort is routinely calculated. These methods were applied to the study of dark matter \cite{kaplan2010atomic}.
However, the authors of the present work are deeply convinced that the question of choosing the correct expression for the cross-section of atomic processes in cosmological structures is not clear enough. For example in the famous paper \cite{zeldovich1978concentration} devoted to density of monopoles in the Universe the authors used the classical approach for the recombination calculations.

Rate of recombination depends on its cross-section. The Kramers formula \eqref{eq1} is typical of kinetic plasma calculations. Its semiclassical expression describes a single-photon recombination for a hydrogen-like atom\cite{kramers1923london}
\begin{equation}
\sigma_{\rm q}\left(n\right)=\frac{32\pi}{3\sqrt{3}}\alpha^3a^2_0\frac{\hbar\omega^2_0}{E\omega n^3},
\label{eq1}
\end{equation} 
where $\hbar\omega_0$ is the energy of ground state, $n$ is the principle number of a bound state, $\omega$ is the frequency of emitted photon, $\alpha$ is the fine-structure constant and $a_0$ is the Bohr radius. 

The approximate summing (coming up to integration) of \eqref{eq1} leads to the following expression
\begin{equation}
\sigma_{\rm q}=\frac{32\pi}{3\sqrt{3}}\alpha r^2_0Z^2\left(\frac{c}{v}\right)^2{\rm ln}\frac{Zc\alpha}{v},
\label{eq2}
\end{equation} 
where $Z$ is a charge of ion, $r_0=\frac{e^2}{m_ec^2}
\sim\alpha$ is the classical electron radius and $v$ is initial velocity.

An another formula for the recombination cross-section was derived by Yelutin \cite{elutin}
\begin{equation}
\sigma_{\rm cl}=\pi\left(4\pi\right)^{\frac{2}{5}}r^2_0Z^{\frac{8}{5}}\left(\frac{c}{v}\right)^{\frac{14}{5}}.
\label{eq3}
\end{equation}
The expression \eqref{eq3} was obtained in terms of the classical mechanics. A single electron is considered as moving from infinity losing its energy due to dipole radiation. When electron's energy becomes zero it comes to bound state.

Expressions \eqref{eq2} and \eqref{eq3} have different conditions of applicability. Formula \eqref{eq2} is valid when
\begin{equation}
Ze^2\gg \hbar v.
\label{eq4}
\end{equation}
That is $v\ll \alpha$ in natural unites ($\hbar=c=1$) and $Z=1$.
It is a common condition for semiclassical approximation in scattering theory \cite{LQM}.

For the classical cross-section we have
\begin{equation}
Z^4\left(\frac{c}{v}\right)^2\gg \alpha^{-5},
\label{eq5}
\end{equation}
i.e. $v\ll \alpha^{5/2}$ in natural units with $Z=1$.

However, if electron's speed satisfies \eqref{eq5}, one will point out dramatic discrepancies. Firstly, the cross-sections have different initial speed dependencies. Secondly, the expression for the classical cross section is several orders of magnitude larger than the quantum. Finally, expressions have different orders of fine-structure constant.

Derivation of an accurate quantum expression for the many-photon recombination is sophisticated problem. The semiclassical consideration of a stimulated  
bremsstrahlung is presented in \cite{berson1981semiclassical}. This paper shows that every partial cross-section depends on its own photon frequency. Establishing a relation between the number of photons and their energies is very complex issue.
Well-known description of quantum single-photon processes is presented in \cite{berestetskii1982quantum}. Although, cross-sections of considered reactions have an analytical form, its derivation is quite cumbersome. The bremsstrahlung cross-section is expressed in terms of the complex hyper-geometric series. Additional photons make establishing correspondence between quantum expression and \eqref{eq3} almost impossible.

To sum it up, the investigation of dark matter led us to the problem on  scattering theory. How to obtain an expression for recombination cross-section when low energy electron emits infinitely many photons? 
Unfortunately, classical monographs devoted to atomic physics do not contain the solution  \cite{sobel2016introduction,BeteSolpiter}.

Here we list some considerations on this topic which do not give solution of the issue. One of the argument, bases on an action, is taken from our previous work \cite{Belotsky_2016}. We just want to collect together something existing for thought just to attract attention to this issue.

\section{
Correspondence between classical expression and Kramers formula}

Now we want to find the situation when two expressions would coincide. In order to do this we have to guess energy loss of an electron. One can notice that if $\omega$ in the denominater of the formula \eqref{eq1} will be changed to E it can coincide with \eqref{eq3}. It will be shown below.

We are considering the electron moving from infinity losint it's energy. Just before the last this iteration(the photon emission) it is possible to use Kramers formula for one-photon recombination. In order to achive the coincidence we will establish the following relation
\begin{equation}
E^2=\tilde{E}\hbar\tilde{\omega}
\label{eq6}
\end{equation} where $\tilde{E}$ is the total energy of the electron before coming to the bound state and $\tilde{\omega}$ is the last emitted photon.

Also the following relations express energy coversation
\begin{equation}
\tilde{E}=\hbar\tilde{\omega}-\frac{\hbar\omega_0}{n^2}
\label{eq7}
\end{equation}
\begin{equation}
\tilde{E}=E-\Delta E
\label{eq8}
\end{equation} where $\Delta E$ is the total energy loss of the electron before last emitting of a photon.

After substitution of \eqref{eq7} and \eqref{eq8} in \eqref{eq6} one will obtain the quadratic equation for $\Delta E$. The solution for $\Delta E$ has the following form
\begin{equation}
\Delta E=E+\frac{\hbar\omega_0}{2n^2}\pm\frac{1}{2}\sqrt{4E^2+\left(\frac{\hbar\omega_0}{n^2}\right)}
\label{eq9}
\end{equation}
It is necessary to choose the sign $-$ to satisfy the energy conservation law.

The next step is summation over new partial cross sections $\tilde{\sigma}_{\rm q} \left(n\right)=\frac{32\pi}{3\sqrt{3}}\alpha^3a^2_0\frac{\hbar\omega^2_0}{E^2n^3}$ starting with number k that we assume bigger than unit to be in the framework of semiclassical limit.
\begin{eqnarray}
\tilde{\sigma}_{\rm q}=\sum_{n=k}^{\infty}\tilde{\sigma}_{\rm q} \left(n\right)=\frac{16\pi Z^4}{3\sqrt{3}}\frac{\alpha^3r^2_0}{k^2}\left(\frac{c}{v}\right)^4
\label{eq10}
\end{eqnarray}

After comparsion of \eqref{eq3} and \eqref{eq10} it is easy to obtain the result for k
\begin{equation}
k=\alpha^{\frac{3}{2}}\left(\frac{c}{v}\right)^{\frac{3}{5}}Z^{\frac{6}{5}}
\label{eq11}
\end{equation} 

Assume $k\gg 1$ and we will obtain
\begin{equation}
Z^4\left(\frac{c}{v}\right)^2\gg \alpha^{-5}
\label{eq12}
\end{equation}
This condition reproduce \eqref{eq5}. Thus, this approach let one point out the connection between two expression: Kramers single-photon cross-section 
can be integrated into the process of bremsstrahlung.

To clarify the physical situation we will give the following reasoning. Slow electron moving from infinity loses it's energy because of bremsstrahlung. In order to understand the correspondence between classical expression \eqref{eq3} and Kramers formula \eqref{eq2} it is necessary to do some manipulations. Firstly, one can notice that when electron initial speed satisfies \eqref{eq5}, energy of a single emitted photon have to be 
relatively large. Electron's energy must be spent on coming to bound state (it has negative energy). Secondly, we rightly assume that if electron overcomes a great distance is not influenced by any other external factors, it will emit many photons. Finally, we establish total energy loss of electron before coming to bound state. Summing all new partial cross sections and comrasion of obtained formula and \eqref{eq3} lets one reproduce original condition \eqref{eq5}.

\section{Estimation of action}

Evaluating of an electron's action also leads to \eqref{eq5}. We will consider an electron moving in the external Coulomb-like field. In order to simplify the calculation charge of ion $Z$ is put equal to unit
\begin{equation}
S=\int_{t_1}^{t_2}\left(\frac{mv^2}{2}+\frac{e^2}{r}\right)dt
\label{eq13}
\end{equation} where $S$ is the action of the electron. 

In the region of interest kinetic energy is proportional to potential
\begin{equation}
\frac{mv^2}{2}\sim\frac{e^2}{r}
\label{eq14}
\end{equation}
This immediately implies $v\sim\sqrt{\frac{2e^2}{mr}}$.
\begin{equation}
S\sim\int_{r_1}^{r_2}2\frac{e^2dr}{rv}\sim\sqrt{me^2}(\sqrt{r_2}-\sqrt{r_1})
\label{eq15}
\end{equation}
Here $r_1$ corresponds to radius of coming to bound state and $r_2$ is the same value with adding the distance, which electron needs to cover for losing most of it's it's initial energy (see \cite{Belotsky_2016}). 
Eventually, if one \textbf{requires }$S\ll\hbar$ 
and obtains
\begin{equation}
\frac{v}{c}\ll\alpha^{\frac{5}{2}}.
\label{eq16}
\end{equation}
Obviously, this condition is in agreement with \eqref{eq5}.
\section{Conclusion}

Calculations originally connected with an estimation of dark matter particles in the Universe generated problem on collision theory. Dark matter is considered as self-interacting according to the law of Coulomb. It immediately implies that darkly charged particles will recombine intensively. Rate of recombination is proportional to the cross-section of its process. Dark matter is considered to have very low energy, what is naturally realised in the Universe (CMB has temperature 3 K, non-relativistic DM should have much lower temperature), what possibly accounts for applicability of classical approximation in a recombination process description. 

This should be understood in the sense of scattering theory. Characteristic speed of particle is lower than atomic speed. This condition is expressed by \eqref{eq4}. In contrast to the Born approximation, semiclassics is applicable here. The problem of the correct expression for the recombination cross-section was discovered. On the one hand, Kramers formula \cite{kramers1923london} is widely used in atomic kinetics. On the other hand, classical expression \cite{elutin} describing the electron capture does not correlate with Kramers cross-section. Of course, 
this discrepancy is connected with the fact that these sections relate to different reactions. Expression describes a single-photon recombination. Formula \eqref{eq3} relates to the process with emitting of infinitely many photons. An attempt to link these two approaches reproduced initial applicability condition \eqref{eq5} for \eqref{eq3}. Moreover, the estimation of the electron's action also leads to the same inequality \eqref{eq16}.

In order to completely solve this problem, it is necessary to obtain expressions for  cross-section of infinitely many photon recombination.
This is a rather ambitious task, but without this the question remains open.

\section*{Acknowledgements}
The work of K.B. was supported by the Ministry of Science and Higher Education of the Russian Federation
by project No 0723-2020-0040 ``Fundamental problems of cosmic rays and dark matter''.
E.E. is also supported by fund Basis, project № 18-1-5-89-1.

Also, we would like to thank A.Barabanov, M.Faifman, V.Lensky, S.Rubin, M.Skorokhvatov for interest to this work and discussions.

\bibliographystyle{unsrt}
\bibliography{bibliography/bibl}
\end{document}